\documentclass[twocolumn, serif]{wiley-article}
\makeatletter
\renewcommand{\@biblabel}[1]{#1.}
\makeatother

\usepackage{graphicx}
\usepackage{float}
\usepackage{stfloats}
\usepackage{booktabs}
\usepackage{multirow}
\usepackage{siunitx}
\usepackage{gensymb}
\usepackage{lineno}
\usepackage{hyperref}
\usepackage{cleveref}
\usepackage{soul, color, xcolor}
\usepackage[numbers,sort&compress]{natbib}
\usepackage{caption}
\usepackage{textcomp,mathcomp}
\usepackage[none]{hyphenat}
\usepackage[markup=default,authormarkup=none]{changes}

\definecolor{red}{HTML}{F44336}
\definecolor{green}{HTML}{4CAF50}
\definecolor{blue}{HTML}{2196F3}
\hypersetup{
colorlinks  = true, 
linkcolor   = blue,
urlcolor    = blue,
citecolor   = blue
}
\crefname{figure}{Figure}{Figure}
\crefname{table}{Table}{Table}
\definechangesauthor[name={Myself}, color=green]{my}

\newcommand{\cit}[1]{\textsuperscript{\cite{#1}}}

\newcommand{\fig}[1]{\textcolor{blue}{\cref{#1}}}
\newcommand{\tab}[1]{\textcolor{blue}{\cref{#1}}}

\newcommand{\dcw}{\textwidth}

\papertype{RESEARCH ARTICLE}
\title{Synergetic Enhancement on Bulk and Grain Boundary Ionic Conduction of Mg Doped High-Entropy NASICON-Type Solid Electrolyte for Solid-State Na$^+$ Batteries by Spray Flame Synthesis}

\author[1]{Tianyi Wu}
\author[2\authfn{1}]{Yiyang Zhang}
\author[1]{Zhu Fang}
\author[1]{Shuting Lei}
\author[3]{Xing Jin}
\author[1]{Shuiqing Li}

\affil[1]{Key Laboratory for Thermal Science and Power Engineering of Ministry of Education, Department of Energy and Power Engineering, Tsinghua University, Beijing, 100084, China}
\affil[2]{Key Laboratory of Advanced Reactor Engineering and Safety of Ministry of Education, Institute of Nuclear and New Energy Technology, Tsinghua University, Beijing, 100084, China}
\affil[3]{Research Center for Gas-phase Synthesis of Functional Nanomaterials, Wuzhen Laboratory, Jiaxing, 314500, China}

\corraddress{\authfn{1}Yiyang Zhang}
\corremail{\textcolor{blue}{zhangyiyang@mail.tsinghua.edu.cn}}

\fundinginfo{National Natural Science Foundation of China,\\ Grant No. \textcolor{blue}{52130606} and \textcolor{blue}{52322608};\\
China Postdoctoral Science Foundation,\\ Grant No. \textcolor{blue}{2023M741894} and \textcolor{blue}{2024T170456};\\
Space Application System of China Manned Space Program}

\runningauthor{Wu et al.}

\begin{document}

\begin{frontmatter}
\maketitle

\begin{abstract}
\setlength{\baselineskip}{0.6\baselineskip}
\textsf{\normalsize All-solid-state sodium batteries represent a promising next-generation energy storage technology, owing to cost-effectiveness and enhanced safety. Among solid electrolytes for solid-state sodium batteries, NASICON-structured Na$\mathsf{_3}$Zr$\mathsf{_2}$Si$\mathsf{_2}$PO$\mathsf{_{12}}$ with complex elemental composition, has emerged as a predominant candidate. However, its widespread implementation remains limited by suboptimal ionic conductivity in both bulk and grain boundary regions. In this study, we demonstrate a novel approach utilizing swirling spray flame synthesis to produce Mg-doped NASICON solid electrolyte nanoparticles. This method facilitates efficient doping and homogeneous mixing for scalable production, resulting in core-shell non-NASICON structures with nano-scale high-entropy mixing characteristics. Notably, the atomic migration distances achieved by flame synthesis are significantly reduced compared to conventional solid-state reactions, thereby enabling reactive sintering to preserve high sinterability of nanoscale particles during subsequent post-treatment processes. High-temperature sintering yields dense NASICON-structured solid electrolytes. Among doping concentrations, Mg$\mathsf{_{0.25}}$NZSP exhibits an optimal ionic conductivity of 1.91 mS/cm at room temperature (25\textcelsius) and an activation energy of 0.200 eV. The enhancement mechanism can be attributed to incorporation into the NASICON phase and formation of a secondary phase by Mg doping. The low-melting-point secondary phase significantly improves grain boundary contact to enhance grain boundary conductivity. The process achieves simultaneous enhancement of both bulk and grain boundary conduction through a single-step procedure instead of complex processes. Comparative analysis of sintering temperatures and ionic conductivities among NASICON solid electrolytes synthesized via different methods demonstrates flame-synthesized nanoparticles offer superior performance and reduced post-treatment costs, owing to their exceptional nano-scale sinterability and uniform elemental distribution.}

\keywords{\setlength{\baselineskip}{0.6\baselineskip}\textsf{\normalsize Swirling spray flame synthesis, Mg-doped NASICON solid electrolyte, Grain boundary, Nano-scale high-entropy mixing, All-solid-state sodium batteries}}
\end{abstract}
\end{frontmatter}

% ========================================================================================
% ===================================== Introduction =====================================
\section{Introduction}
\label{sec:Introduction}
The development of next-generation battery systems has garnered significant attention in recent years, driven by the increasing demand for enhanced energy density and safety characteristics\cit{dunn2011electrical, kim2012current, zhu2021lithium}. The substitution of conventional organic liquid electrolytes with solid electrolytes (SEs) in all-solid-state batteries (ASSBs) presents a transformative approach which not only eliminates critical safety concerns such as leakage, flammability, and explosion risks but also addresses current intrinsic limitations in energy density\cit{manthiram2017lithium, gao2018promises, chen2019approaching}. Such innovation demonstrates substantial potential for widespread implementation across various applications, including electric vehicles, energy conversion and storage systems, and aerospace propulsion technologies\cit{pan2013room, hueso2013high}. Furthermore, sodium-ion batteries (SIBs) have emerged as a compelling alternative to conventional lithium-ion batteries (LIBs), offering distinct advantages in terms of economic viability for sodium's abundant natural reserves, widespread geographical distribution, and substantially lower resource costs. Additionally, sodium's reduced reactivity compared to lithium contributes to enhanced safety characteristics\cit{deng2018sodium, zhao2020rational}. Consequently, all-solid-state sodium-ion batteries are widely recognized as promising candidates for next-generation energy storage systems\cit{zhao2018solid}.\\
\indent Developing solid-state sodium batteries fundamentally relies on the advancement of solid electrolytes\cit{liu2024inorganic, wang2019development}, among which Na$_3$Zr$_2$Si$_2$PO$_{12}$ (NZSP) named as Na super ionic conductor (NASICON), has attracted mostly investigation for its exceptional ionic conductivity, wide electrochemical stability window, and remarkable resistance to moisture exposure, thermal fluctuations, and chemical degradation\cit{boilot1987crystal, li2022nasicon}. Nevertheless, the ionic conductivity of this oxide-based solid electrolyte remains a limiting factor in its practical implementation for solid-state batteries\cit{ahmad2023recent}. To enhance Na-ion conductivity, elemental doping has proven to be an effective strategy\cit{yang2020ultrastable, jolley2015improving, wang2023enhanced, zhang2018effect, yu2019sodium}. And the NASICON-type structure offers multiple accessible sites for various dopant elements, providing extensive opportunities for conductivity optimization through strategic doping design\cit{rizvi2024recent, ouyang2021synthetic, wang2023design, zeng2022high}.\\
\indent As a polycrystalline ceramic material, NASICON solid electrolytes encounter ionic conduction resistance through two distinct mechanisms: bulk resistance within individual grains and grain boundary resistance at the interfaces between grains, both of which affect Na$^+$ ion transport\cit{zhao2019solid, li2022insight}. Elemental doping typically induces lattice distortion, creating new migration pathways and modifying carrier concentrations, thereby enhancing Na$^+$ mobility within NASICON grains and improving bulk conductivity. However, this approach generally has minimal impact on grain boundary conductivity\cit{rizvi2024recent}. Enhancement of grain boundary conductivity has been achieved through liquid-phase sintering, wherein low-melting-point additives form a liquid phase during the sintering process, facilitating improved intergranular contact\cit{oh2019composite, noi2018liquid}. Consequently, achieving superior overall ionic conductivity necessitates simultaneous enhancement of both bulk and grain boundary conduction mechanisms. Nevertheless, this optimization presents significant challenges due to the complex interplay of multiple elements and phases, coupled with intricate processing requirements.\\
\indent To synthesize complex multi-element oxide nanoparticles, flame synthesis presents distinct advantages over conventional solid-state reaction (SSR) and liquid-phase sol-gel processes. As a gas-phase synthesis technique, flame synthesis offers unique capabilities including facile doping procedures, flexible process control, and nano-scale high-entropy mixing. Additionally, this method enables scalable production of nano-sized particles through a continuous, environmentally sustainable process\cit{li2016flame, schulz2019gas, teoh2010flame, kelesidis2017flame, haddad2018sno2}. In recent years, flame synthesis has become an effective approach for large-scale production of functional materials across diverse applications, including battery technology\cit{wu2024spray, wu2022rapid}, catalysis\cit{koirala2016synthesis, strobel2004flame}, ceramics\cit{kammler2001flame, pratsinis1998flame}, and optical materials\cit{lei2025spray, lei2024two, zhao2007synthesis}. These characteristics make flame synthesis particularly well-suited for the production of doped NZSP nanoparticles.\\
\indent In this work, we demonstrate the synthesis of Mg-doped NZSP nanoparticles using a scalable swirling spray flame synthesis method, employing Mg element as an economically viable dopant. Comprehensive morphological and structural characterization reveals that these five-element particles, exhibiting nano-scale high-entropy mixing, are formed through a bottom-up gas-phase process, significantly facilitating subsequent post-treatment procedures. The electrical properties of NZSP solid electrolytes with varying Mg dopant concentrations are systematically investigated to elucidate the distinct effects of Mg incorporation on both bulk and grain boundary conduction mechanisms. Furthermore, we present a comparative analysis of sintering requirements and corresponding ionic conductivities among Na$_3$Zr$_2$Si$_2$PO$_{12}$-based NASICON structural solid electrolytes synthesized via different methodologies, highlighting the significant advantages of gas-phase flame synthesis.

% ========================================================================================
% ========================================================================================

% ========================================================================================
% ================================ Results and Discussion ================================
\section{Results and Discussion}
\label{sec:Results and Discussion}

% ========================================== 2.1 =========================================
\subsection{Structure and Morphology}
\label{ssec:Structure and Morphology}
\begin{figure*}[t]
\centering
\includegraphics[width=\dcw]{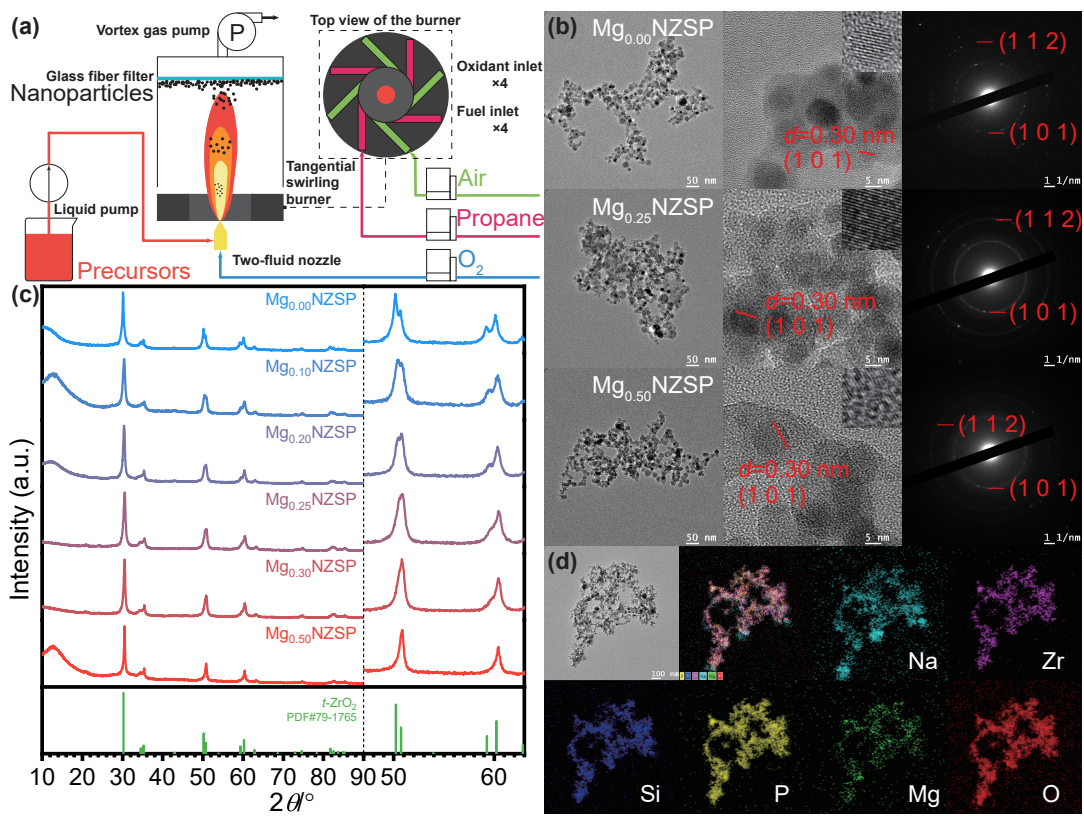}
\captionsetup{justification=justified, singlelinecheck=false}
\caption{Synthesis and characterization of Mg-doped NZSP nanoparticles. a) Schematic illustration of the swirling spray flame synthesis system for Mg-doped NZSP nanoparticle production. b) TEM images and corresponding SAED patterns of as-synthesized Mg$_{0.00}$NZSP, Mg$_{0.25}$NZSP, and Mg$_{0.50}$NZSP nanoparticles. c) XRD patterns of as-synthesized Mg$_x$NZSP nanoparticles ($x$=0-0.5). d) TEM micrograph and corresponding EDS elemental mapping of as-synthesized Mg$_{0.25}$NZSP nanoparticles.}
\label{fig:Fig1}
\end{figure*}

Transmission electron microscopy (TEM) analysis of as-synthesized Mg$_0$NZSP, Mg$_{0.25}$NZSP, and Mg$_{0.5}$NZSP nanoparticles is presented in \fig{fig:Fig1}b. The Mg$_{x}$NZSP nanoparticles with uniform diameters of approximately 10 nm are successfully synthesized by swirling spray flame synthesis system. These particles exhibit a distinctive core-shell structure, comprising crystalline cores encapsulated by non-crystalline shells. The crystalline cores display a lattice spacing of 0.30 nm, corresponding to the (1 0 1) plane of tetragonal ZrO$_2$ (\textit{t}-ZrO$_2$). This structural configuration arises from the differential gas-to-particle temperature characteristics of Zr and other constituent elements. During the cooling process at flame field, ZrO$_2$ initially nucleates and grows, followed by the subsequent solidification and deposition of other oxides on the surface of existing crystalline \textit{t}-ZrO$_2$ cores, forming amorphous outer layers\cit{wu2024spray}. Selected area electron diffraction (SEAD) patterns of the as-synthesized Mg$_0$NZSP, Mg$_{0.25}$NZSP, and Mg$_{0.5}$NZSP nanoparticles (\fig{fig:Fig1}b) further confirm the crystalline of the inner cores. Analysis of the diffraction patterns reveals the presence of (1 0 1) and (1 0 2) planes of \textit{t}-ZrO$_2$, corroborating the identification of the inner crystalline phase as tetragonal ZrO$_2$. Measurement of the other as-synthesized nanoparticle samples is also supplied in \textcolor{blue}{Figure S1}.\\
\indent X-ray diffraction (XRD) analysis of the as-synthesized Mg$_x$NZSP ($x$=0-0.5) samples is presented in \fig{fig:Fig1}c with tetragonal structure of ZrO$_2$ (ICSD Database, PDF\#79-1765) shown as a reference pattern. The crystalline structure is exclusively comprised of Zr-based phases forming the inner crystalline cores, while oxides of Na, Si, and P constitute the amorphous outer shells. This structural configuration results from the distinct gas-to-particle temperature profiles, leading to spatiotemporally heterogeneous nucleation and growth processes within the flame field. Mg incorporation into the ZrO$_2$ structural cores manifests as systematic lattice distortions in the XRD patterns. Detailed analysis of the 2$\theta$=47$\degree$-63$\degree$ region in \fig{fig:Fig1}c reveals notable structural evolution with increasing Mg content. At approximately 2$\theta$=50$\degree$, Mg$_{0}$NZSP exhibits two obvious different individual diffraction peaks corresponding to (1 1 2) and (2 0 0) crystal planes. Progressive Mg doping induces peak coalescence and a systematic shift toward higher angles, indicating lattice contraction due to the substitution of host ions with smaller-radius Mg$^{2+}$ ions. Similar peak evolution is observed for the (1 0 3) and (2 1 1) planes near 60$\degree$. Detailed peak-shift analysis is offered in \textcolor{blue}{Figure S2}. These XRD patterns corroborate the TEM observations, confirming the core-shell structure comprising \textit{t}-ZrO$_2$ crystalline cores surrounded by amorphous shells, rather than exhibiting complete atomic-level mixing of all constituent elements\cit{wu2024spray}.\\
\indent Energy-dispersive X-ray spectroscopy (EDS) mapping of an additional region of as-synthesized Mg$_{0.25}$NZSP nanoparticles is presented in \fig{fig:Fig1}d, illustrating the spatial distribution of constituent elements Na, Zr, Si, P, Mg, and O. While \fig{fig:Fig1}b demonstrates distinct phase separation between Zr and other elements rather than complete atomic-level mixing, the flame synthesis process nevertheless achieves relatively uniform elemental distribution at the nanoscale. This nanoscale homogeneity facilitates the formation of the desired NASICON structure during subsequent sintering processes, ultimately enabling superior electrochemical performance.
% ------------------------------------------- Fig.1 --------------------------------------
% ----------------------------------------------------------------------------------------

% ========================================================================================

% ========================================== 2.2 =========================================
\subsection{Fabrication and Sinterability}
\label{ssec:Fabrication and Sinterability}

\begin{figure*}[t]
\centering
\includegraphics[width=\dcw]{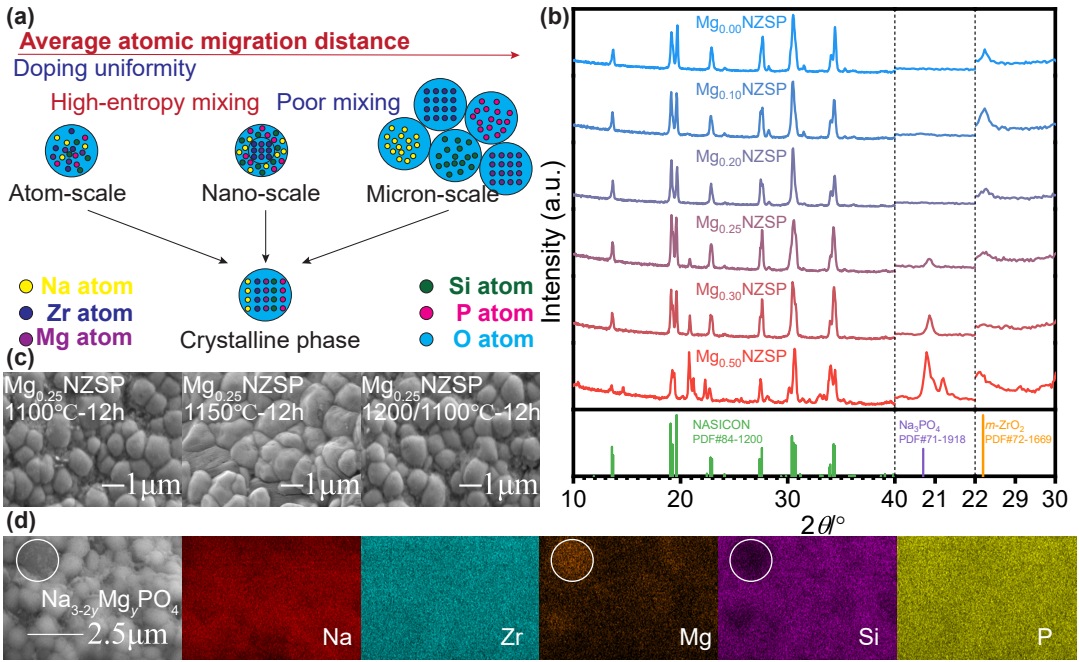}
\captionsetup{justification=justified, singlelinecheck=false}
\caption{Phase formation mechanisms and microstructural characterization of sintered Mg-doped NZSP electrolytes. a) Schematic illustration depicting elemental mixing scales and corresponding NASICON phase formation mechanisms in Mg$_{x}$NZSP particles. b) XRD patterns of two-step sintered (1200/1100$\celsius$ for 12h) Mg$_x$NZSP electrolyte pellets ($x$=0-0.5). c) SEM cross-sectional micrographs of Mg$_{0.25}$NZSP electrolyte pellets processed under various sintering conditions. d) SEM micrograph and corresponding energy-dispersive X-ray spectroscopy (EDS) elemental mapping of Mg$_{0.25}$NZSP electrolyte sintered at 1200/1100$\celsius$ for 12h.}
\label{fig:Fig2}
\end{figure*}

The structural and morphological characterization of as-synthesized Mg$_{x}$NZSP nanoparticles reveals uniform nano-scale particles produced via spray flame synthesis, though lacking the desired NASICON phase. These core-shell structured nanopowders, in their non-NASICON configuration, exhibit poor electrochemical performance, necessitating post-treatment for phase transformation. While high-temperature annealing can induce the formation of monoclinic NASICON structure, prolonged exposure to elevated temperatures typically results in particle growth, adversely affecting sinterability. However, the distinctive advantage of flame synthesis for achieving uniform mixing at the nanoscale presents novel opportunities for optimizing post-processing protocols and phase formation strategies.\\
\indent \fig{fig:Fig2}a illustrates three distinct levels of elemental mixing and their corresponding phase formation mechanisms. While flame synthesis ideally aims to achieve uniform atomic-level distribution of dopant elements, the distinct physicochemical properties of these elements within the flame field result in core-shell structures with nanoscale mixing. Nevertheless, this configuration maintains advantages of high-entropy mixing, characterized by significantly reduced average atomic migration distances during phase formation compared to the micron-scale mixing typical of conventional solid-state reaction methods. Consequently, raw Mg$_x$NZSP nanoparticles can undergo reactive sintering, combining densification and NASICON phase formation while maintaining both high sinterability and superior electrical properties.\\
\indent  Leveraging the advantages of flame synthesis, we employ reactive sintering methodology combined with two-step sintering, wherein electrolyte pellet densification occurs concurrently with NASICON microstructure formation. The flame-synthesized nanoparticles, characterized by high-entropy mixing and enhanced sinterability, enable rapid elemental diffusion at significantly reduced temperatures which is approximately 100$\sim$150$\celsius$ lower than those required for conventional solid-state reaction methods in NASICON-structural NZSP-based systems\cit{samiee2017divalent, li2022insight, go2021investigation, yang2018nasicon}. Based on these advantages, we establish optimal sintering protocols: reactive sintering at 1100$\celsius$ and 1150$\celsius$, and two-step sintering at 1200/1100$\celsius$, successfully producing dense Mg$_x$NZSP electrolyte pellets.\\
\indent XRD analysis of sintered electrolyte pellets confirms the formation of the desired NASICON structure. \fig{fig:Fig2}b presents XRD patterns of Mg$_x$NZSP pellets sintered at 1200/1100$\celsius$ for 12h. For lower Mg dopant concentrations ($x$=0, 0.1, and 0.2), the patterns primarily exhibit monoclinic NASICON phase peaks (ICSD Database, PDF\#84-1200) with minor contributions from monoclinic ZrO$_2$ (ICSD Database, PDF\#72-1669), consistent with undoped samples. The expanded 2$\theta$=28$\degree$-30$\degree$ region is also shown here for comparison. As Mg content increases beyond $x$=0.2$\sim$0.25, the NASICON structure exhibits limited Mg solubility, resulting in the formation of a secondary phase identified as Mg-doped Na$_3$PO$_4$ (ICSD Database, PDF\#71-1918), significantly evident in samples with $x$=0.25, 0.3, and 0.5. The emergence of this secondary phase is clearly demonstrated in the expanded 2$\theta$=20$\degree$-22$\degree$ region. The systematic right shift of the diffraction peak near 2$\theta$=21$\degree$ relative to pure Na$_3$PO$_4$ indicates the formation of Na$_{3-2y}$Mg$_y$PO$_4$, attributed to lattice contraction along the (1 1 1) plane due to the smaller ionic radius of Mg$^{2+}$ compared to Na$^+$. Detailed analysis is shown in \textcolor{blue}{Figure S3}. The structural evolution exhibits distinct behavior depending on Mg concentration: direct incorporation into the crystal structure at lower concentrations, versus secondary phase formation at higher concentrations. This compositional dependence significantly influences the Na$^+$ conduction properties of the NASICON solid electrolyte, as detailed in subsequent analyses.\\
\indent The sintered electrolytes exhibit high densification, with Mg$_0$NZSP achieving relative densities of 95.1\%, 96.0\%, and 96.4\% (theoretical density $\rho_0$=3.27 $\rm{g/cm^3}$\cit{oh2019composite}) when sintered at 1100$\celsius$-12h, 1150$\celsius$-12h, and 1200/1100$\celsius$-12h, respectively. Both elevated temperatures and two-step sintering protocols effectively enhance densification of NZSP electrolytes\cit{wu2024spray}. Scanning electron microscopy (SEM) analysis of sintered Mg$_x$NZSP electrolyte pellet cross-sections in \fig{fig:Fig2}c reveals dense microstructures with minimal porosity across all sintering protocols. Notably, samples processed at 1100$\celsius$-12h and 1200/1100$\celsius$-12h exhibit refined grain sizes, attributed to lower holding temperatures that suppress grain growth. EDS mapping of an additional region of Mg$_{0.25}$NZSP sintered at 1200/1100°C-12h in \fig{fig:Fig2}d reveals distinctive elemental distribution patterns. Zr distribution analysis is omitted due to spectral overlap with P. While Na and P exhibit uniform distribution, Mg and Si show heterogeneous spatial arrangements. This heterogeneity stems from the absence of Si in the Na$_{3-2y}$Mg$_y$PO$_4$ secondary phase and differential Mg solubility between phases. The limited Mg incorporation in the NASICON phase results in preferential Mg enrichment within the Na$_{3-2y}$Mg$_y$PO$_4$ phase, corroborating the XRD findings presented in \fig{fig:Fig2}b. This phase distribution significantly influences the electrical characteristics of the electrolyte further.

% ------------------------------------------- Fig.2 --------------------------------------
% ----------------------------------------------------------------------------------------

% ========================================================================================

% ========================================== 2.3 =========================================
\subsection{Electrical Conductivity}
\label{ssec:Electrical Conductivity}
\begin{figure*}[t]
\centering
\includegraphics[width=\dcw]{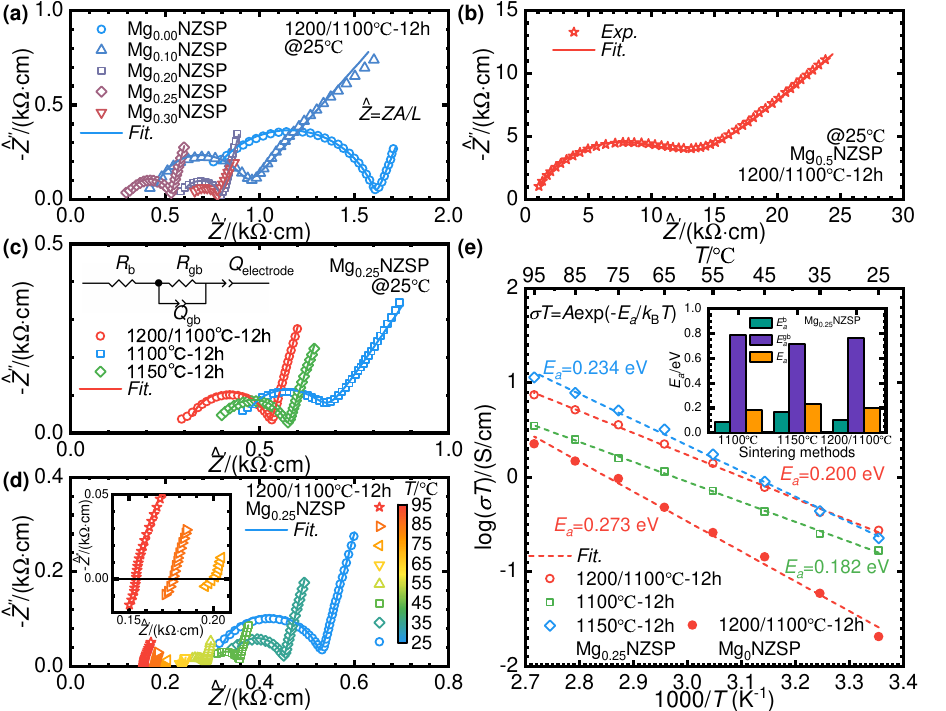}
\captionsetup{justification=justified, singlelinecheck=false}
\caption{Electrochemical characterization of Mg-doped NZSP electrolytes. Normalized Nyquist plots at 25$\celsius$ for a) Mg$_{x}$NZSP, $x$=0-0.3, and b) Mg$_{0.5}$NZSP electrolyte symmetrical cells processed by two-step sintering (1200/1100$\celsius$-12h). c) Normalized impedance spectra at 25$\celsius$ for Mg$_{0.25}$NZSP electrolytes prepared using various sintering protocols. d) Temperature-dependent normalized impedance spectra (25-95$\celsius$) for Mg$_{0.25}$NZSP electrolyte sintered at 1200/1100$\celsius$-12h. (e) Arrhenius plots comparing ionic conductivity of Mg$_{0.25}$NZSP electrolytes processed under different sintering conditions with undoped Mg$_{0}$NZSP (1200/1100$\celsius$-12h). Inset: Comparative analysis of bulk, grain boundary, and total activation energies for Mg$_{0.25}$NZSP samples across different sintering protocols.}
\label{fig:Fig3}
\end{figure*}

Following morphological and structural characterization, the ionic conductivity of sintered Mg$_x$NZSP electrolyte pellets is evaluated using electrochemical impedance spectroscopy (EIS). Symmetrical cells are constructed using ion-sputtered Au electrodes into an Au\textbar Mg$_x$NZSP\textbar Au configuration. \fig{fig:Fig3} presents the normalized Nyquist plots, where impedance values are expressed as $\hat{Z}=ZA/L$ to eliminate geometric dependencies. In this normalization, $\hat{Z}$ represents the normalized impedance, $Z$ denotes the measured impedance, and $A$ and $L$ correspond to the cross-sectional area and thickness of the disk-shaped electrolyte pellets, respectively.\\
\indent \fig{fig:Fig3}ab present normalized Nyquist plots and corresponding fitting results for Mg$_x$NZSP electrolyte pellets sintered at 1200/1100$\celsius$-12h, demonstrating composition-dependent conductivity at room temperature (25$\celsius$). The ionic conductivity exhibits a non-monotonic relationship with Mg content, reaching a maximum of 1.91 mS/cm for Mg$_{0.25}$NZSP, which is a significant enhancement compared to the undoped sample (0.62 mS/cm). While Mg doping effectively promotes electrical conductivity, optimal performance requires precise compositional control, achievable through flame synthesis. Excess Mg incorporation, as observed in Mg$_{0.3}$NZSP, deteriorates conductive performance. \\
\indent Panel c displays normalized Nyquist plots with equivalent circuit fitting for Mg$_{0.25}$NZSP samples prepared by different sintering protocols at 25$\celsius$. The equivalent circuit comprises bulk ($R_{\rm{b}}$) and grain boundary ($R_{\rm{gb}}$) resistances, with constant phase elements for grain boundary ($Q_{\rm{gb}}$) and Au electrode blocking ($Q_{\rm{electrode}}$). The two-step reactive sintering protocol (1200/1100$\celsius$-12h) yields superior performance, attributed to enhanced densification. While higher temperature reactive sintering (1150$\celsius$-12h) outperforms lower temperature processing (1100$\celsius$-12h), the two-step protocol achieves superior properties at lower maximum temperature, demonstrating advantages for NASICON-type Na$^+$ electrolytes\cit{wu2024spray}. Additionally, the two-step process maintains refined grain size despite high density, enhancing mechanical properties crucial for solid-state sodium battery applications. Individual Nyquist plots at 25$\celsius$ of the samples with various Mg concentrations sintered by 1200/1100$\celsius$-12h, 1100$\celsius$-12h, and 1150$\celsius$-12h are also supplied in \textcolor{blue}{Figure S4}, \textcolor{blue}{Figure S5}, and \textcolor{blue}{Figure S6}, respectively. \\
\indent Temperature-dependent impedance measurements of Mg$_{0.25}$NZSP sintered by 1200/1100$\celsius$-12h from 25-95°C are presented in \fig{fig:Fig3}d. Ionic conduction exhibits strong temperature dependence, with enhanced Na$^+$ diffusion across grains and grain boundaries at elevated temperatures. At 75-95$\celsius$ (inset in \fig{fig:Fig3}d), the significantly reduced resistance manifests as incomplete semicircles in Nyquist plots, where the real axis intersection represents total resistance. \fig{fig:Fig3}e presents the Arrhenius relationships between ionic conductivity $\sigma$ and absolute temperature $T$, expressed as $\sigma T=A\exp(-E_a/k_{\rm{B}}T)$, where $A$ represents the pre-exponential factor, $k_{\rm{B}}=1.38\times 10^{-23} \rm{J/K}$ is the Boltzmann constant, and $E_a$ denotes the activation energy characterizing the sensibility of electro-conductivity with operating temperature. Detailed data is supplied in \textcolor{blue}{Table S1}. Comparative analysis of Mg$_{0.25}$NZSP samples processed at 1100$\celsius$-12h, 1150$\celsius$-12h, and 1200/1100$\celsius$-12h against undoped Mg$_{0}$NZSP (1200/1100$\celsius$-12h) confirms the efficacy of Mg doping in enhancing ionic conductivity across practical operating temperatures. Doped electrolytes demonstrate 1-2 fold higher conductivity than undoped samples over 25-95$\celsius$, with activation energies consistent with literature values\cit{noi2018liquid}. Mg$_{0.25}$NZSP processed at 1200/1100$\celsius$-12h and 1150$\celsius$-12h exhibit optimal ionic conductivities, with activation energies of 0.200 eV and 0.234 eV, respectively. This results in temperature-dependent performance crossover: 1200/1100$\celsius$-12h samples demonstrate superior conductivity below 35$\celsius$, while 1150$\celsius$-12h samples excel at higher temperatures. To elucidate processing-structure-property relationships, bulk ($E_a^{\rm{b}}$) and grain boundary ($E_a^{\rm{gb}}$) activation energy for Mg$_{0.25}$NZSP are determined using $\sigma_i T=A_i\exp(-E_a^i/k_{\rm{B}}T)$, $i=\rm{g}, \rm{gb}$ in the inset of \fig{fig:Fig3}e. Samples processed at 1100$\celsius$-12h and 1200/1100$\celsius$-12h, characterized by similar grain sizes due to identical holding temperatures, exhibit comparable activation energies. In contrast, samples processed at 1150$\celsius$-12h display elevated bulk activation energy but slightly reduced grain boundary activation energy, attributed to enhanced grain growth at higher holding temperatures. These microstructural differences manifest as distinct temperature-dependent conductivity behaviors across processing conditions.

% ------------------------------------------- Fig.3 --------------------------------------
% ----------------------------------------------------------------------------------------

\begin{figure*}[t]
\centering
\includegraphics[width=\dcw]{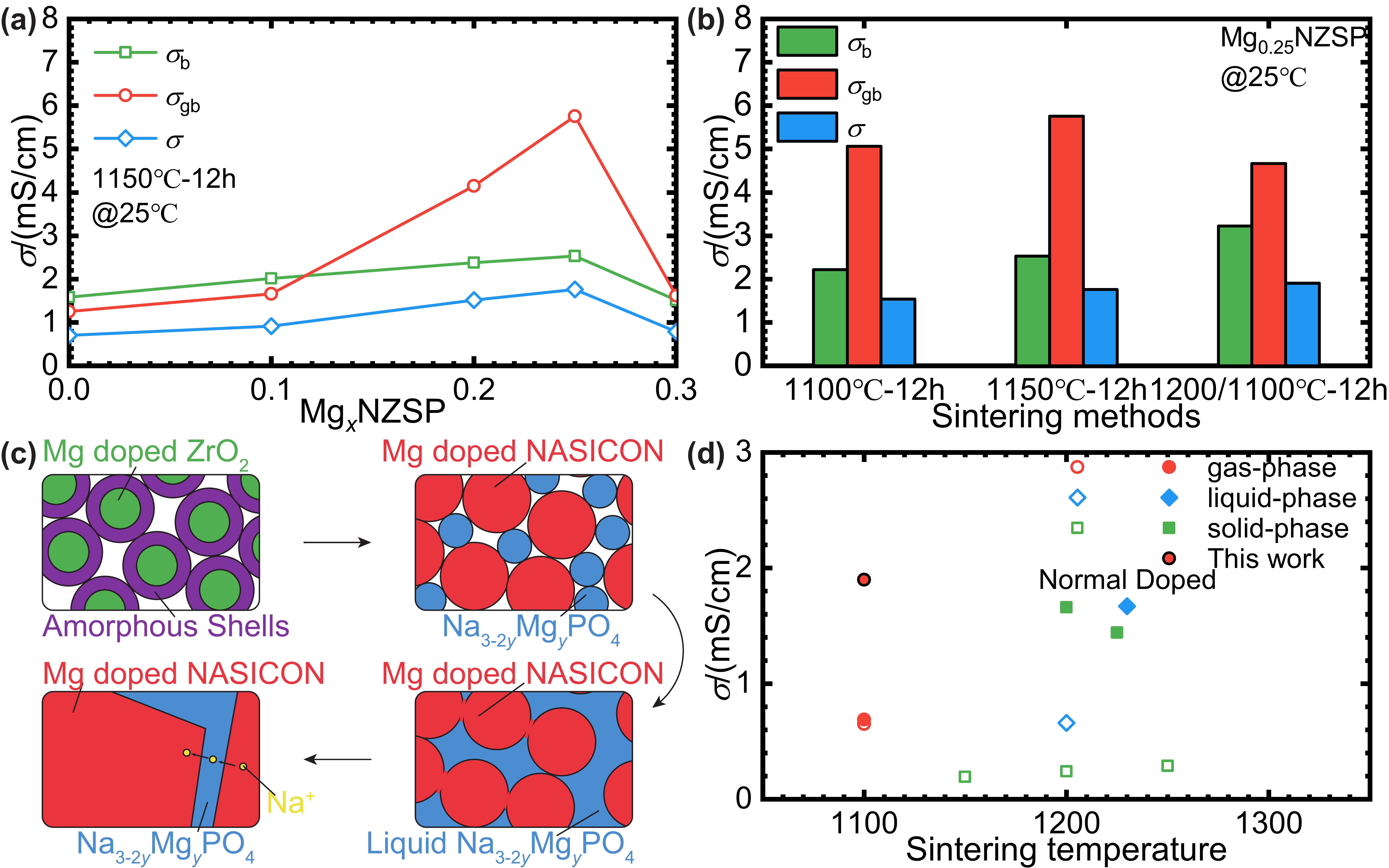}
\captionsetup{justification=justified, singlelinecheck=false}
\caption{Conductivity analysis and mechanistic insights of Mg-doped NASICON electrolytes. (a) Room temperature (25$\celsius$) bulk ($\sigma_{\rm{b}}$), grain boundary ($\sigma_{\rm{gb}}$), and total ($\sigma$) ionic conductivities as functions of Mg content for samples sintered at 1150$\celsius$-12h. b) Comparative analysis of bulk, grain boundary, and total conductivities for Mg$_{0.25}$NZSP processed under various sintering conditions. c) Schematic illustration of microstructural evolution and Na$^+$ transport enhancement mechanism in Mg-doped NASICON electrolytes, highlighting secondary phase formation at grain boundaries. d) Performance comparison of Na$_3$Zr$_2$Si$_2$PO$_{12}$-based NASICON electrolytes: sintering temperature versus electrical conductivity for pristine and doped compositions synthesized via different processing routes.}
\label{fig:Fig4}
\end{figure*}

To elucidate the influence of Mg doping and sintering protocols on ionic transport, we analyze bulk ($\sigma_{\rm{b}}$), grain boundary ($\sigma_{\rm{gb}}$), and total electrical conductivities ($\sigma$=$\sigma_{\rm{b}}$$\sigma_{\rm{gb}}$/($\sigma_{\rm{b}}$+$\sigma_{\rm{gb}}$)), reflecting distinct Na$^+$ transfer mechanisms within grains and across grain boundaries. \fig{fig:Fig4}a and \textcolor{blue}{Table S2} illustrates the compositional dependence of these conductivities at 25$\celsius$ for samples sintered at 1150$\celsius$-12h. All conductivity components exhibit non-monotonic behavior with Mg content, with $\sigma_{\rm{gb}}$ showing pronounced variation near Mg$_{0.25}$NZSP, coinciding with Na$_{3-2y}$Mg$_y$PO$_4$ secondary phase formation. This secondary phase enhances grain boundary conductivity through improved intergrain contact. However, excessive Mg doping leads to performance degradation due to excessive secondary phase formation, indicating that Mg doping primarily influences NASICON conductivity through grain boundary modification. \fig{fig:Fig4}b compares the impact of different sintering protocols on conductivity components in Mg$_{0.25}$NZSP at room temperature. Reactive sintering at 1150$\celsius$-12h yields higher conductivities compared to 1100$\celsius$-12h processing, attributed to enhanced densification. The two-step sintering protocol (1200/1100$\celsius$-12h) achieves maximal bulk conductivity through optimal densification, although grain boundary conductivity slightly decreases due to refined grain size and increased boundary density. However, in Mg-doped systems, grain boundary conductivity remains significantly higher than bulk conductivity, making the latter the limiting factor according to $\sigma$=$\sigma_{\rm{b}}$$\sigma_{\rm{gb}}$/($\sigma_{\rm{b}}$+$\sigma_{\rm{gb}})$. Consequently, the two-step sintering approach effectively enhances total ionic conductivity through bulk conductivity optimization while maintaining favorable mechanical properties. These results demonstrate that combined reactive and two-step sintering protocols optimize NASICON electrolyte performance through simultaneous enhancement of ionic conductivity and mechanical properties.\\
\indent The preceding analysis demonstrates that enhanced grain boundary conductivity, facilitated by Na$_{3-2y}$Mg$_y$PO$_4$ secondary phase formation, primarily drives the improvement in total ionic conductivity. \fig{fig:Fig4}c presents a proposed transformation mechanism elucidating this process. Initially, core-shell structured nanoparticles, prior to NASICON phase formation, are compacted into disk-shaped pellets. During thermal processing, as thermodynamic conditions for NASICON phase formation are achieved, the limited Mg solubility in the NASICON structure leads to concurrent Na$_{3-2y}$Mg$_y$PO$_4$ phase separation. The nano-scale high-entropy mixing achieved through flame synthesis ensures that both phases originate from individual particles of nanometer dimensions, resulting in uniform and interpenetrating phase distribution. As sintering progresses at elevated temperatures, Mg-doped NASICON grains coalesce and grow while the Na$_{3-2y}$Mg$_y$PO$_4$ phase transitions to a liquid state. The incorporation of Mg into the crystal lattice advantageously modifies the thermal characteristics of Na$_3$PO$_4$: elevating its volatilization temperature to prevent high-temperature losses while maintaining a sufficiently low melting point to enable liquid phase formation. This liquid phase effectively infiltrates the intergranular regions between NASICON grains. Upon cooling, the solidified Na$_{3-2y}$Mg$_y$PO$_4$ forms conductive layers at grain boundaries, establishing enhanced pathways for Na$^+$ transport. This microstructural evolution significantly improves grain boundary conductivity, consequently enhancing total ionic conductivity through the formation of optimized ion-conducting networks at interfaces.\\
\indent The demonstrated enhancement in ionic conductivity through Mg doping and subsequent Na$_{3-2y}$Mg$_y$PO$_4$ secondary phase formation highlights the effectiveness of grain boundary engineering. Furthermore, the nano-scale high-entropy mixing achieved through gas-phase flame synthesis enables reactive sintering, facilitating simultaneous densification and phase formation while maintaining high sinterability at reduced processing temperatures. \fig{fig:Fig4}d presents a comparative analysis of sintering temperatures and electrical conductivities for Na$_3$Zr$_2$Si$_2$PO$_{12}$-based NASICON electrolytes synthesized via gas-phase\cit{wu2024spray, liu2019substituted}, liquid-phase\cit{rao2021influence, lu2019high}, and solid-phase\cit{ruan2019optimization, go2021investigation, rao2021influence, li2022insight, chen2018influence} methods. For conventional Na$_3$Zr$_2$Si$_2$PO$_{12}$ compositions\cit{wu2024spray, rao2021influence, ruan2019optimization, go2021investigation}, gas-phase and liquid-phase synthesis achieve comparable ionic conductivities, though gas-phase processing requires markedly lower sintering temperatures. Solid-state reaction products exhibit inferior performance despite higher processing temperatures. While elemental doping generally enhances performance\cit{liu2019substituted, lu2019high, li2022insight, chen2018influence}, liquid-phase and solid-phase approaches still necessitate elevated sintering temperatures for adequate densification. The present work, employing gas-phase flame synthesis with Mg doping, uniquely combines enhanced sinterability at reduced processing temperatures with superior ionic conductivity. Detailed data is listed in \textcolor{blue}{Table S3}. The distinctive advantages of gas-phase flame synthesis, nanoscale particle production, facile dopant incorporation, and high-entropy mixing, enable the fabrication of high-performance electrolytes with reduced processing costs. These attributes establish a promising pathway for scalable production of NASICON-based solid electrolytes.

% ------------------------------------------- Fig.4 -------------------------------------
% ----------------------------------------------------------------------------------------

% ========================================================================================

% ========================================================================================
% ===================================== Conclusions ======================================
\section{Conclusions}
\label{sec:Conclusions}
In this study, we leverage the inherent doping flexibility of a high-throughput swirling spray flame synthesis system (>1 kg/h production capacity) to produce Mg-doped NZSP electrolyte nanoparticles with approximately 10 nm diameter. The as-synthesized particles exhibit a core-shell architecture with nano-scale high-entropy mixing, resulting in reduced atomic migration distances for crystalline phase formation. This unique structural characteristic enables the implementation of reactive sintering, where simultaneous phase formation and densification preserve the enhanced sinterability of nanoscale particles while requiring reduced processing temperatures.\\
\indent Post-sintering characterization reveals dense NASICON-phase electrolyte pellets with composition-dependent microstructural evolution. Increasing Mg content promotes the formation of Na$_{3-2y}$Mg$_y$PO$_4$ secondary phase, which effectively fills intergranular regions, enhancing grain boundary contact and ionic transport. The optimized composition, Mg$_{0.25}$NZSP, processed via combined reactive and two-step sintering, achieves superior Na$^+$ ionic conductivity of 1.91 mS/cm at room temperature (25$\celsius$) with an activation energy of 0.200 eV, representing a 1-2 fold enhancement over undoped compositions. Comparative analysis of NZSP electrolytes synthesized through various methods demonstrates that gas-phase flame synthesis uniquely combines enhanced performance with reduced processing temperatures. These attributes, coupled with scalable production capability, position this approach as a promising pathway for commercial solid-state sodium battery applications.\\
\indent This study represents initial investigations into swirling spray flame synthesis (SFS) of NZSP electrolytes, focusing specifically on single-element Mg doping in $\rm{Na_3Zr_2Si_2PO_{12}}$. The potential for enhanced performance through multi-element co-doping strategies, particularly targeting Zr, Si, and P sites to create high-entropy structures\cit{ouyang2021synthetic}, remains largely unexplored. Previous studies suggest that such compositional complexity can significantly enhance the electrochemical properties of NASICON-structured electrolytes\cit{wang2023design, zeng2022high}. Given that facile dopant incorporation represents a key advantage of flame synthesis, future work will explore multi-element doping strategies via swirling SFS to develop higher-performance NASICON conductors.
% ========================================================================================
% ========================================================================================

% ========================================================================================
% ================================= Experimental Methods =================================
\section{Experimental Methods}
\label{sec:Experimental Methods}

% ========================================== 4.1 =========================================
\subsection{Swirling Spray Flame Synthesis System}
\label{ssec:Swirling Spray Flame Synthesis System}
Mg-doped NASICON nanoparticles ($\rm{Na}_{3+2\textit{x}}\rm{Zr}_{2-\textit{x}}\rm{Mg}_{\textit{x}}\rm{Si}_2\rm{PO}_{12}$) are synthesized using a swirling spray flame synthesis system, as illustrated in \fig{fig:Fig1}a. The process involves atomization of liquid precursor solutions via an atomizing gas stream into a swirling spray flame burner, where ignition by tangential swirling flames generates the reactive spray flame environment. Within the high-temperature flame field, nanoparticle formation proceeds through sequential processes including chemical reaction, nucleation, collision, coagulation, and surface growth etc.\cit{wu2023clustering}, followed by particle collection for subsequent processing.\\
\indent Precursor solutions and atomizing $\rm{O}_2$ are continuously introduced into a two-fluid nozzle positioned beneath the tangential swirling burner at flow rates of 0.6 L/h and 16 L/min, respectively. The high-velocity gas flow shears the liquid phase into microscale droplets. The tangential swirling burner features eight narrow slots (1 mm width $\times$ 15 mm height) at the cavity base, enabling separate but adjacent injection of gaseous fuel (Propane, $\rm{CH_3CH_2CH_3}$, 1.4 L/min) and oxidant (Air, 32 L/min). This configuration promotes efficient mixing through high-velocity flow, generating stable swirling flames for precursor ignition. The synthesized nanoparticles are separated from the gas stream via glass fiber filtration driven by a vortex gas pump. The collected nanopowders undergo subsequent compression and sintering for electrochemical characterization. The photograph of lab-scale system is shown in \textcolor{blue}{Figure S7}.
% ========================================================================================

% ========================================== 4.2 =========================================
\subsection{Preparation of Precursor Solutions}
\label{ssec:Preparation of Precursor Solutions}
Precursor solutions for Mg-doped NASICON nanoparticle synthesis were prepared using commercially available metal acetates and nitrates. The synthesis employs sodium acetate ($\rm{NaCH_3COO}$, 99\%), zirconium nitrate pentahydrate ($\rm{Zr(NO_3)_4\cdot 5H_2O}$, 99.9\%), magnesium acetate tetrahydrate ($\rm{Mg(CH_3COO)_2\cdot 4H_2O}$, 99\%), hexamethyldisiloxane ($\rm{((CH_3)_3Si)_2O}$, 99.9\%), and triethyl phosphate ($\rm{(CH_3CH_2O)_3PO}$, 99.9\%) dissolved in ethyl alcohol ($\rm{CH_3CH_2OH}$) at 0.5 mol/L total concentration. Small additions of propionic acid ($\rm{CH_3CH_2COOH}$) and deionized water ($\rm{H_2O}$) facilitated complete dissolution, ensuring uniform elemental distribution without visible precipitates. 2-ethylhexanoic acid (EHA, $\rm{CH_3(CH_2)_3CH(C_2H_5)COOH}$) is incorporated stoichiometrically with metal-containing inorganic precursors to promote homogeneous nanopowder formation\cit{strobel2011effect,wei2019investigating}. To compensate for volatilization losses, Na and P precursors were added in 10\% excess\cit{ruan2019optimization}. Leveraging the compositional flexibility of flame synthesis, a series of $\rm{Na}_{3+2\textit{x}}\rm{Zr}_{2-\textit{x}}\rm{Mg}_{\textit{x}}\rm{Si}_2\rm{PO}_{12}$ compositions ($x$=0, 0.1, 0.2, 0.25, 0.3, and 0.5, denoted as Mg$_x$NZSP) are investigated. Detailed precursor solution parameters are presented in \tab{tab:Tab1}. While the current laboratory-scale production achieves approximately 20 g/h, successful scale-up to over 1 kg/h has been demonstrated through optimization of combustion parameters and collection efficiency in an industrial-scale device.

% ------------------------------------------- Tab.1 --------------------------------------
\begin{table*}[htbp]
    \centering
    \caption{Compositional parameters of precursor solutions for Mg-doped $\rm{Na}_{3+2\textit{x}}\rm{Zr}_{2-\textit{x}}\rm{Mg}_{\textit{x}}\rm{Si}_2\rm{PO}_{12}$ (Mg$_x$NZSP) synthesis.}
    \label{tab:Tab1}
    \begin{tabular*}{\textwidth}{@{\extracolsep{\fill}} lrrrrrr}
        \toprule
        \textbf{Precursor}                  & \multicolumn{6}{c}{\textbf{Concentration} (mol/L)} \\
        \midrule
                                            & \textbf{Mg$_{0}$NZSP}     & \textbf{Mg$_{0.1}$NZSP}   & \textbf{Mg$_{0.2}$NZSP}   & \textbf{Mg$_{0.25}$NZSP}  & \textbf{Mg$_{0.3}$NZSP}   & \textbf{Mg$_{0.5}$NZSP}                       \\
        $\rm{NaCH_3COO}$                    & 0.196                     & 0.204                     & 0.212                     & 0.215                     & 0.219                     & 0.232                                         \\
        $\rm{Zr(NO_3)_4\cdot 5H_2O}$        & 0.119                     & 0.110                     & 0.102                     & 0.098                     & 0.094                     & 0.079                                         \\
        $\rm{Mg(CH_3COO)_2\cdot 4H_2O}$     & 0                         & 0.006                     & 0.011                     & 0.014                     & 0.017                     & 0.026                                         \\
        $\rm{((CH_3)_3Si)_2O}$              & 0.119                     & 0.116                     & 0.113                     & 0.112                     & 0.110                     & 0.105                                         \\
        $\rm{(CH_3CH_2O)_3PO}$              & 0.066                     & 0.064                     & 0.062                     & 0.062                     & 0.061                     & 0.058                                         \\
        EHA                                 & 0.673                     & 0.657                     & 0.641                     & 0.634                     & 0.627                     & 0.600                                         \\
        \bottomrule
    \end{tabular*}
\end{table*}
% ----------------------------------------------------------------------------------------

% ========================================== 4.3 =========================================
\subsection{Cell Fabrication}
\label{ssec:Cell Fabrication}
The electrochemical characterization of synthesized materials employs symmetrical cells with Au\textbar Mg$_x$NZSP\textbar Au configuration. Raw nanopowders were initially compressed at 200 MPa into disk-shaped pellets ($\phi$=10 mm diameter). Subsequent sintering protocols employ reactive sintering at either 1100$\celsius$ or 1150$\celsius$ for 12h (denoted as 1100$\celsius$-12h and 1150$\celsius$-12h, respectively). The thermal treatment protocol maintained a heating/cooling rate of 5$\celsius$/min below 950$\celsius$ and 1$\celsius$/min above 950$\celsius$. Additionally, a combined reactive and two-step sintering approach\cit{chen2000sintering,kim2022interface,huang2018two} is implemented, wherein samples are heated to 1200$\celsius$ followed by immediate reduction to 1100$\celsius$ for a 12h isothermal hold (denoted as 1200/1100$\celsius$-12h), maintaining identical heating/cooling rates. The sintered Mg$_x$NZSP electrolyte pellets are subsequently coated with Au blocking electrodes via ion sputtering on both faces to form symmetrical cells for electrochemical measurements.
% ========================================================================================

% ========================================== 4.4 =========================================
\subsection{Material Characterization}
\label{ssec:Material Characterization}
Morphological and compositional analyses of Mg-doped NZSP nanoparticles are performed using transmission electron microscopy (TEM, JEM 2010, JEOL Ltd., 120 kV) and X-ray energy dispersive spectrometry (EDS, B5-U92). Crystallinity is confirmed through selected area electron diffraction (SAED). Surface microstructure and elemental distribution of sintered electrolyte pellets are examined using high-resolution scanning electron microscopy (SEM, JSM7401F, JEOL Ltd.) operated at 3.0 kV coupled with EDS. Specific surface area (SSA) measurements of Mg$_x$NZSP nanoparticles are conducted using a gas physisorption and chemisorption analyzer (ASAP 2460, Micromeritics) following the Brunauer-Emmett-Teller (BET) method. Assuming monodisperse, non-aggregated spheres, the BET-equivalent diameter is calculated using $d_{\rm{BET}}=6/(\rho_0 \cdot \rm{SSA})$, where $\rho_0$ represents the theoretical density of the nanocrystalline electrolyte. Phase analysis is performed using X-ray diffraction (XRD, SmartLab 9 kW, Rigaku) with filtered Cu K$\rm{\alpha}$ radiation ($\lambda=1.54\ \si{\angstrom}, 40\ \rm{kV}, 150\ \rm{mA}$). Diffraction patterns are collected over $2\theta=10\degree-90\degree$ at a scan rate of 5$\degree$/min with 0.04$\degree$ resolution.
% ========================================================================================

% ========================================== 4.5 =========================================
\subsection{Electrochemical Tests}
\label{ssec:Electrochemical Tests}
The ionic conductivities of the fabricated symmetrical cells were measured using open-circuit electrochemical impedance spectroscopy (EIS) with a electrochemical workstation (BioLogic, SP-300 Potentiostat). Measurements were conducted at temperatures ranging from 25$\celsius$ to 95$\celsius$ in 10$\celsius$ intervals, with impedance frequencies $f=$ spanning from 7.0 MHz to 0.1 Hz. The theoretical impedance profile of an electrolyte symmetrical cell comprises two semicircles and one sloping linear tail occurring at different frequency regimes. The first semicircle in the high-frequency regime represents Na$^+$ ion migration through grain lattice sites via a hopping mechanism, with its intercept on the real axis corresponding to the bulk resistance $R_{\rm{b}}$. At medium frequencies, the ion transfer process across grain boundaries, accompanied by electric double layer formation due to Na$^+$ concentration and depletion at opposite sides of the grain boundary, manifests as the second semicircle in the Nyquist plot. The difference between the intercepts of this semicircle defines the grain boundary resistance $R_{\rm{gb}}$. The sloping linear tail in the low-frequency region is attributed to Na$^+$ transfer resistance at the electrolyte by Au blocking electrode interface. Due to limitations in the frequency measurement range, the complete impedance profile may not be obtained, and the first semicircle in the high-frequency regime may be absent. Equivalent circuit modeling is employed to fit the EIS data and extract values for both bulk resistance $R_{\rm{b}}$ and grain boundary resistance $R_{\rm{gb}}$. The total ohmic resistance is given by $R=R_{\rm{b}}+R_{\rm{gb}}$. The ionic conductivity $\sigma$ is calculated using the relationship $\sigma=L/(RA)$, where $L$ represents the thickness and $A$ denotes the cross-sectional area of the symmetrical cell.
% ========================================================================================

% ========================================================================================
% ========================================================================================

\section*{Supporting Information}
Supporting Information is available from the Wiley Online Library or from the author.

\section*{acknowledgements}
We acknowledge support from the National Natural Science Foundation of China (Grant No. \textcolor{blue}{52130606} and \textcolor{blue}{52322608}), the China Postdoctoral Science Foundation (No. \textcolor{blue}{2023M741894} and \textcolor{blue}{2024T170456}), and the Space Application System of China Manned Space Program. The authors would like to thank Mr. Du Li at Tsinghua university and Ms. Xiaoya Zhang, Ms. Shaohua Zhang, and Mr. Linhuan Zhang at Wuzhen laboratory for fruitful discussions and helps on experiments and characterization.

\section*{conflict of interest}
The authors declare no conflict of interest.

\printendnotes

\bibliographystyle{MSP}

\bibliography{ref}

\end{document}